\documentclass[sigconf, authorversion, nonacm]{acmart}
\usepackage{graphicx,subcaption}
\AtBeginDocument{%
  }

\setcopyright{acmlicensed}
\copyrightyear{2024}
\acmYear{2024}
\acmDOI{XXXXXXX.XXXXXXX}

\acmConference[Conference acronym 'XX]{Make sure to enter the correct
  conference title from your rights confirmation email}{June 03--05,
  2024}{Woodstock, NY}
\acmISBN{978-1-4503-XXXX-X/18/06}

\newcommand{\food}{\emph{/r/Food}}

\newcommand{\xhdr}[1]{\vspace{0.7mm}\noindent{{\bf #1.}}}

\begin{document}

\title{NutriTransform: Estimating Nutritional Information\\From Online Food Posts}

\author{Thorsten Ruprechter}
\affiliation{%
  \institution{Graz University of Technology}
  \country{Austria}
}
\email{th.ruprechter@gmail.com}

\author{Marion Garaus}
\affiliation{%
  \institution{Sigmund Freud Private University}
  \city{Vienna}
  \country{Austria}}
\email{marion.garaus@sfu.ac.at}

\author{Ivo Ponocny}
\affiliation{%
  \institution{Sigmund Freud Private University}
  \city{Vienna}
  \country{Austria}}
\email{ivo.ponocny@sfu.ac.at}

\author{Denis Helic}
\affiliation{%
  \institution{Graz University of Technology}
  \city{Graz}
  \country{Austria}}
\email{dhelic@tugraz.at}

\renewcommand{\shortauthors}{Ruprechter et al.}

\begin{abstract}
Deriving nutritional information from online food posts is challenging, particularly when users do not explicitly log the macro-nutrients of a shared meal.
In this work, we present an efficient and straightforward approach to approximating macro-nutrients based solely on the titles of food posts.
Our method combines a public food database from the U.S. Department of Agriculture with advanced text embedding techniques.
We evaluate the approach on a labeled food dataset, demonstrating its effectiveness, and apply it to over $500\,000$ real-world posts from Reddit's popular /r/food subreddit to uncover trends in food-sharing behavior based on the estimated macro-nutrient content.
Altogether, this work lays a foundation for researchers and practitioners aiming to estimate caloric and nutritional content using only text data.
\end{abstract}

\begin{CCSXML}
<ccs2012>
   <concept>
       <concept_id>10002951.10003317</concept_id>
       <concept_desc>Information systems~Information retrieval</concept_desc>
       <concept_significance>300</concept_significance>
       </concept>
   <concept>
       <concept_id>10003120.10003130.10003233</concept_id>
       <concept_desc>Human-centered computing~Collaborative and social computing systems and tools</concept_desc>
       <concept_significance>500</concept_significance>
       </concept>
   <concept>
       <concept_id>10002951.10003260</concept_id>
       <concept_desc>Information systems~World Wide Web</concept_desc>
       <concept_significance>500</concept_significance>
       </concept>
 </ccs2012>
\end{CCSXML}

\ccsdesc[300]{Information systems~Information retrieval}
\ccsdesc[500]{Human-centered computing~Collaborative and social computing systems and tools}
\ccsdesc[500]{Information systems~World Wide Web}

\keywords{digital traces, macro-nutrient estimation, online food sharing, embedding techniques, food data}


\maketitle

\section{Introduction}
\begin{figure*}[t]
	\centering
	\begin{subfigure}[b]{.30\linewidth}
		\centering
	       \frame{\includegraphics[width=\linewidth]{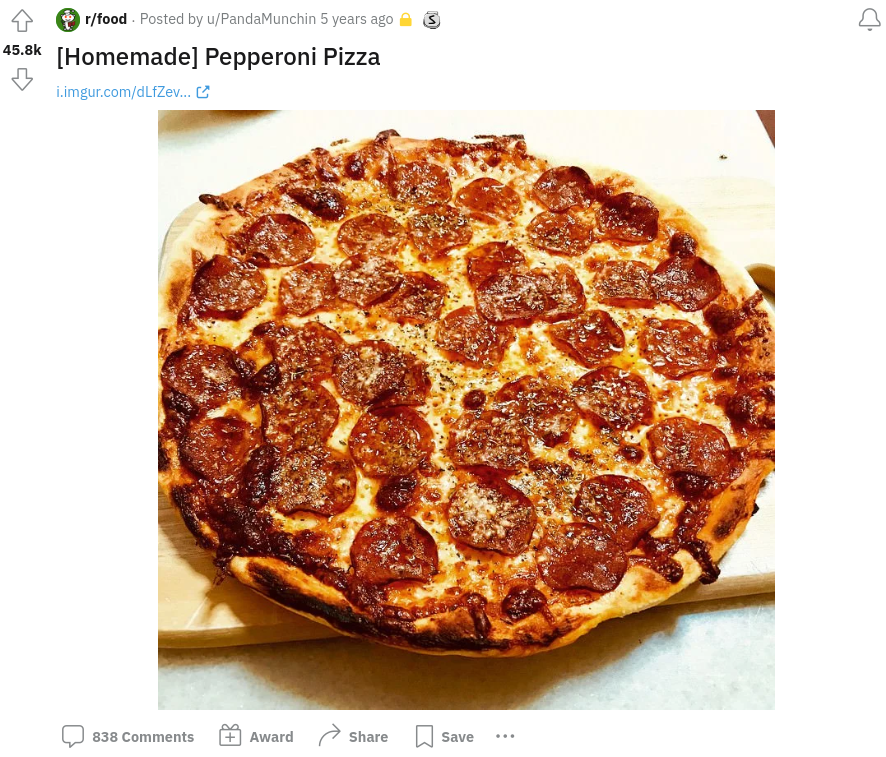}}
	   \caption{Example Submission on \food.}
	\label{fig:food_sub}
	\end{subfigure}
    \hfill
	\begin{subfigure}[b]{.34\linewidth}
		\centering
	       \includegraphics[width=\linewidth]{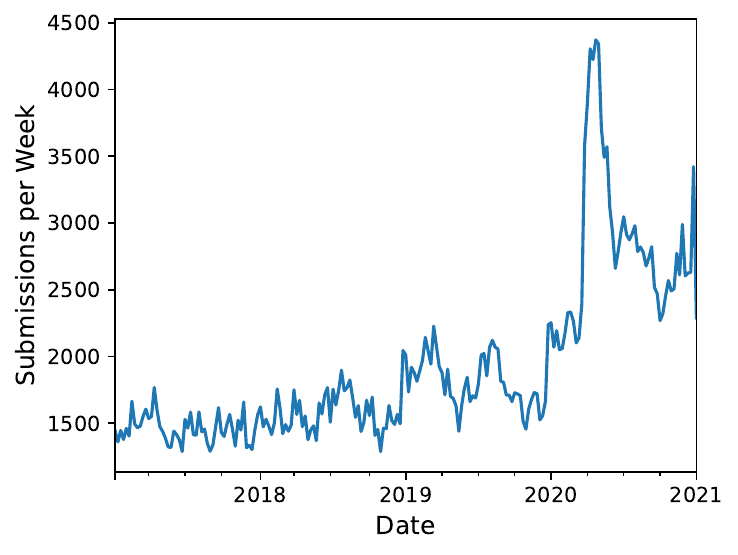}
	   \caption{Number of Posts to \food.}
	\label{fig:food_post_counts}
	\end{subfigure}
 	\begin{subfigure}[b]{.34\linewidth}\includegraphics[width=\linewidth]{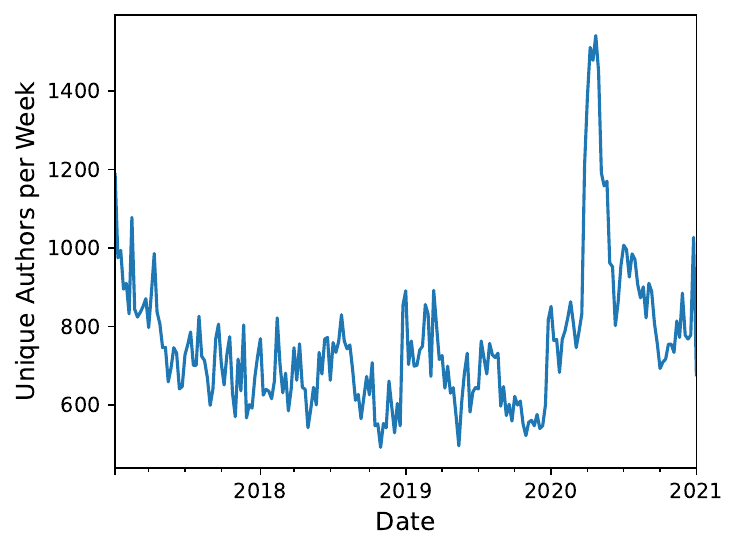}
	   \caption{Number of Unique Authors in \food.}
	\label{fig:food_author_counts}
	\end{subfigure}
    
	\caption{\textbf{Food posts on Reddit.} 
	The /r/food subreddit is one of the largest sub-communities on the popular online discussion platform Reddit. 
	Users share food-related posts within this community, receiving engagement in the form of upvotes (i.e., likes) or comments (Fig.~\ref{fig:food_sub}).
	Since 2017, /r/food has consistently received between $1\,300$ and $2\,300$ posts per week, with activity peaking at over $4\,000$ weekly posts following the onset of the COVID-19 pandemic in March 2020 (Fig.~\ref{fig:food_post_counts}).
	Similarly, the early phases of the pandemic saw the highest number of unique weekly contributors, with over $1\,400$ individual users posting each week (Fig.~\ref{fig:food_author_counts}).
		}
	\label{fig:reddit_post}
\end{figure*}

As our personal lives increasingly move online, tracking online behavior has become a meaningful way to measure human activity and interactions~\cite{lazer2021meaningful}.
Users now generate a wealth of online traces by interacting with posts on social media platforms such as \emph{Twitter/X}~\cite{gligoric2020experts}, by editing and viewing articles on Wikipedia~\cite{ruprechter2023poor}, or by submitting and commenting on posts in online discussion forums such as Reddit~\cite{russo2024stranger}.
These online behavioral traces offer researchers opportunities to analyze users' habits and customs, drawing parallels between their digital presence and offline life.
However, deriving behavioral insights from online activity, often expressed in short texts, is challenging due to the limited information they typically provide. 
Despite these limitations, researchers can use these brief contributions as proxies for estimating details that users do not explicitly share. 
Among such diverse online traces that are often times difficult to interpret, food-related content offers a unique perspective on individual habits and societal trends. 
For instance, in certain forums on the online discussion platform Reddit, users post pictures of food with an image and a brief title (Fig.~\ref{fig:food_sub}).
As users rarely provide detailed ingredient lists for their food posts, however, estimation of more advanced metrics such as macro-nutrient composition or caloric content can prove difficult~\cite{de2016characterizing}.
Nonetheless, approximating such information can yield valuable insights into users' dietary patterns and global eating trends.

Therefore, in this work, we propose an approach to estimate nutritional values based solely on the title of users' food posts.
Using the USDA food database and SentenceTransformer embeddings, we evaluate our method on a labeled dataset and apply it to real-world food posts from Reddit.
To support further research and practical applications, we make our code openly available,\footnote{\url{https://github.com/ruptho/nutritransform}} providing a tool for researchers and practitioners to estimate caloric content from textual data across various use cases.

\begin{figure*}[t]
	\includegraphics[width=\linewidth]{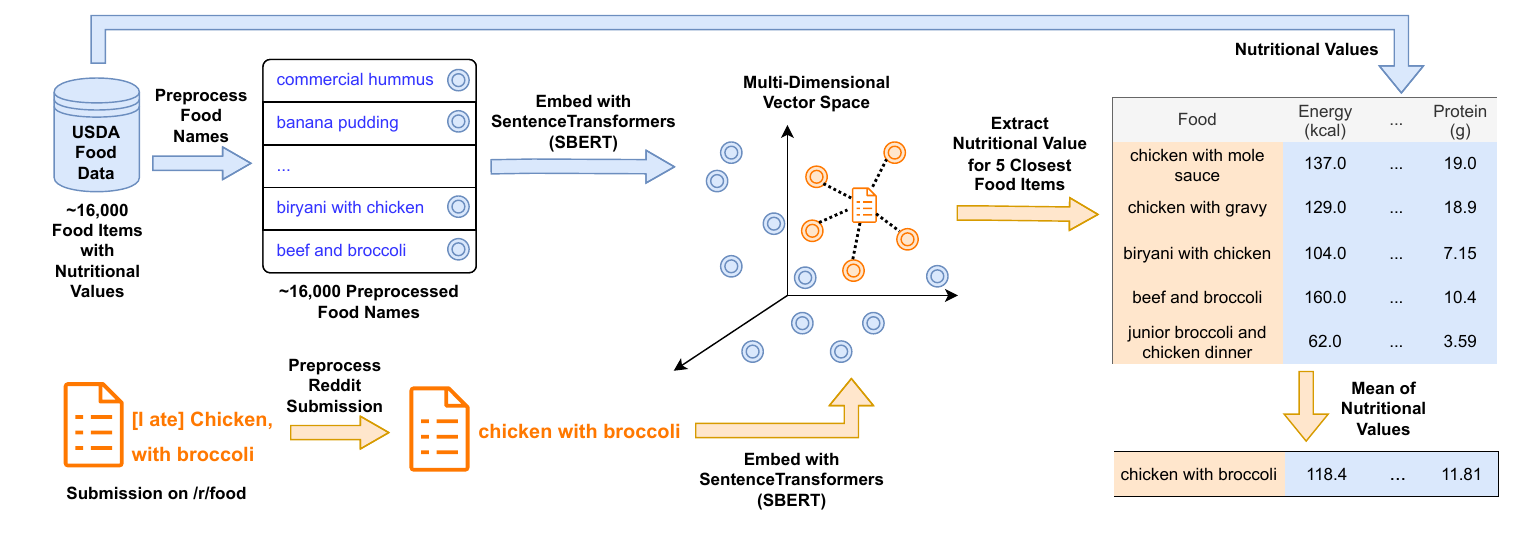}
	\caption{\textbf{Pipeline for estimating nutritional values.}
	We first generate SentenceTransformer embeddings from all foods from the USDA food database.
	Afterwards, when labeling a new food post, such as a submission to /r/food on Reddit, we embed the title of the post and retrieve the $n$ most similar entries that exceed a predefined similarity threshold $t$. 
	Finally, we aggregate the nutritional values of these most-similar items to estimate the macro-nutrients (e.g., calories) for the unlabeled food titles.}
	\label{fig:pipeline}
\end{figure*}

\section{Nutritional estimation}

To estimate nutritional values from food titles, we utilize entries in the USDA food databases~\cite{fukagawa2022usda} and leverage BERT sentence embeddings~\cite{reimers2019sbert} (see Fig.~\ref{fig:pipeline}).
We use the dataset from West et al.~\cite{west2013cookies}, which contains recipes from a website along with extracted nutritional values, as a gold-standard dataset for tuning our estimation approach.

\xhdr{USDA foods database} 
The USDA Foods database is an open resource provided by the U.S. Department of Agriculture, offering various sources of food data~\cite{fukagawa2022usda}. 
We collect foods and their corresponding nutritional values from three USDA data sources: 
Foundation Foods (commodity-derived basic foods), the Food and Nutrient Database for Dietary Studies (FNDDS, referred to as ``Survey foods''), and the National Nutrient Database for Standard Reference (``SR Legacy'').\footnote{USDA FoodData Central: \url{https://fdc.nal.usda.gov/}} 
These datasets are widely used in dietary studies and represent an openly available standard for nutritional information.
Additionally, since a significant portion of Reddit’s user demographic is U.S.-based or originates from other Western countries, this database is generally well-suited to our purposes.
However, we acknowledge that focusing on this dataset may exclude more granular information about ethnic foods from other parts of the world, even though their nutritional value can still be estimated using our transformer-based approach.
%\CITE{Describe that food databases exist \cite{delgado2021food}. Mention USDA \cite{fukagawa2022usda}, FoodDB \cite{durazzo2022food}, EuroFIR~\cite{harrington2019nutrient} as examples. State that we choose USDA because it suits the US-heavy demographic of Reddit and it is a well-established resource in past research [cite]

\xhdr{Preprocessing and embedding food titles}
The food names in the USDA databases often follow a structured, hierarchical format (e.g., ``Mushrooms, portabella, grilled'') .
To improve compatibility with natural language usage, we preprocess these strings by converting the text to lowercase, tokenizing the comma-separated phrases, and reversing the token order to better reflect typical phrasing (e.g., ``grilled portabella mushroom''). 
Furthermore, we remove duplicates and exclude uncooked or raw food items (i.e., food titles containing “raw” or “uncooked”), resulting in our foundational database of $14\,264$ foods.
Finally, we use SentenceTransformers~\cite{reimers2019sbert} with the \texttt{all-mpnet-base-v2} model to generate embeddings for the preprocessed food titles.

\xhdr{Estimating calories}
To estimate the nutritional value of food titles not covered in our database, we leverage the SentenceTransformer embeddings of our foundational database alongside its nutritional information provided by the USDA (Fig.~\ref{fig:pipeline}).
First, we preprocess the target food titles by converting them to lowercase and removing commas before generating their SentenceTransformer embeddings.
Using these embeddings, we identify the \texttt{n} closest USDA food items based on cosine similarity and filter out items with a similarity below a threshold \texttt{t} to exclude potentially unrelated matches.
The nutritional value of the original food title is then estimated by aggregating the nutritional information of the remaining most similar items.
To determine the optimal values for the number of neighbors (\texttt{n}), similarity threshold (\texttt{t}), and aggregation function (\texttt{m})---i.e., whether to combine the values of the retrieved \texttt{n} items using the mean, median, or weighted mean (weighted by their cosine similarity to the target)---we conduct hyperparameter tuning using a labeled dataset from West et al.~\cite{west2013cookies}.

\xhdr{Hyperparameter tuning}
For tuning, we utilize the recipe dataset from West et al.~\cite{west2013cookies}, which includes 8,865 food titles with corresponding recipes and nutritional information. We split the dataset into a training set (80\%, $7,092$ recipe titles) and a test set (20\%, $1,773$ recipe titles). Using our approach, we estimate calories per 100 grams\footnote{We hereafter focus on calories as they combine multiple macronutrient components into a single metric.} for all recipes in the training set (mean = $207.21$ calories, std. dev. = $130.04$).
We test various values for \texttt{n} ($1, 5, 10, 20, 25, 50, 75, 100$), \texttt{t} ($0.0, 0.5, 0.75, 0.9$), and \texttt{m}, and find the best configuration of \texttt{n}=50, \texttt{t}=0.0, and \texttt{m}=\textit{weighted mean}, based on a root mean squared error (RMSE) of $114.76$. This configuration produces an RMSE of $116.78$ on the held-out test set, indicating good generalization.
For comparison, we evaluate the CalorieNinjas API,\footnote{\url{https://calorieninjas.com}} a black-box natural language processing-based calorie estimation tool, on the same test set. 
The API produces an RMSE of $122.71$, higher than our method, further demonstrating the competitiveness of our approach.

\begin{figure*}[t]
	\centering
	\includegraphics[width=\linewidth]{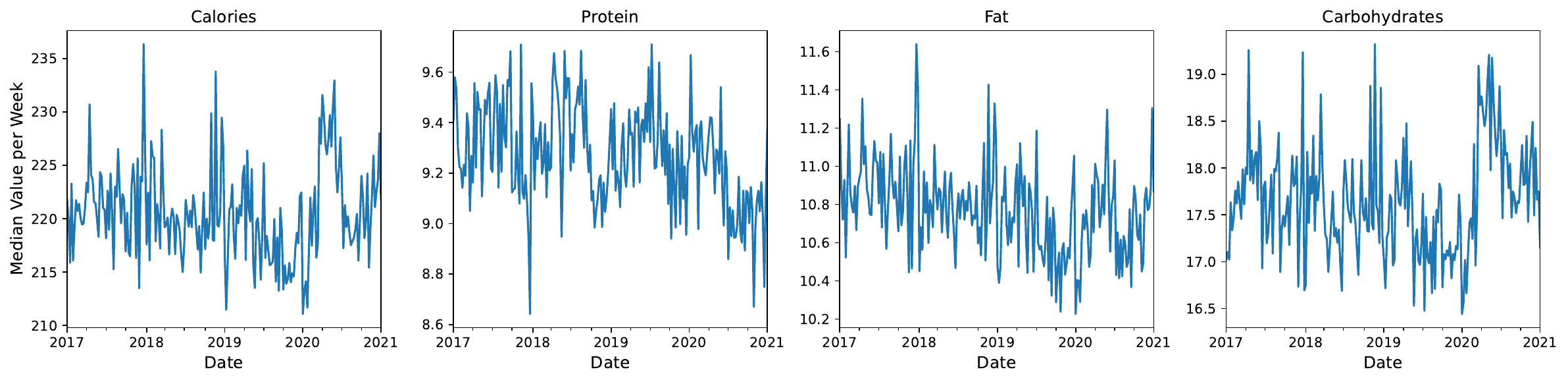}
    \caption{\textbf{Nutritional values of foods shared on Reddit over the years.}
    We visualize weekly medians of food posts on r/food for four nutritional metrics: calories, protein, fat, and carbohydrates per 100 grams. Our investigation reveals general trends, such as higher calorie counts toward the end of most years, as well as a notable plateau from March to June 2020, coinciding with the onset of the COVID-19 pandemic.}
	\label{fig:macro_nutrients}
\end{figure*}

\xhdr{Differences to traditional approaches}
Our approach offers several advantages over traditional natural language processing methods, which are often based on dependency parsing or n-grams.
Most notably, the vocabulary of the submission title does not need to exactly match the food name in the database. For example, even if the database contains an entry for “tomato paste,” the SentenceTransformer embedding for “tomato purée” would still yield valid nutritional information due to their semantic similarity.
Additionally, our method is robust to typos, plural forms, and variations in phrasing (e.g., “broccoli chicken” is similar to “chicken with broccoli”).
Traditional approaches often require exact matches or elaborate title parsing, which still provide only approximate nutritional values due to the brevity of food titles.
Our method effectively addresses these limitations, offering a more flexible and accurate solution.

\section{Food Conversations on Reddit}
To showcase the capabilities of our approach, we now apply our calorie approximation to posts about food shared by real users on the online conversation platform Reddit.

\xhdr{Food posts on Reddit}
Reddit is one of the most popular websites on the Internet. Visitors of Reddit can read and discuss their interests in separate sub-forums (``subreddits'').
In these subreddits, users can contribute content by creating standalone posts (``submissions'') or commenting on existing submissions.
One of the largest subreddit communities is \food\footnote{Note that the \textit{/r/} on Reddit is a prefix indicating subreddits names similar to \@ indicating user names on Twitter.}.
In this subreddit, users submit pictures of food and state its name in the submission title (see Fig.~\ref{fig:food_sub}).
Submissions to \food\ must follow certain rules or otherwise be deleted by manual or automatic moderation.\footnote{Although rules change from time to time, they have been mostly consistent between 2015 and 2021.}
First, 70\% of a submission's title has to specifically describe the food itself, while the submission body should contain only an image.
Additionally, the submission title must contain a tag that indicates whether users cooked the food at home (``[Homemade]''), they ate it at a restaurant (``[I ate]''), or they are a professional chef (``[Pro/Chef]'').
Lastly, \food\ only allows for dishes that users photographed themselves.
For this work, we utilize all submissions made to \food\ between 2017 and 2021.
Consequently, we first crawl $757\,605$  food submissions using the pushshift.io dataset dumps \cite{baumgartner2020pushshift}.
We then ignore submissions that were either deleted, are duplicates (by title, day, and user), or do not have a valid tag in the submission title (either ``[Homemade]'', ``[I ate]'', or ``[Pro/Chef]'').
Our resulting dataset contains $513\,303$ submissions by $157\,577$ unique users (Figs.~\ref{fig:food_post_counts} and~\ref{fig:food_author_counts}).
%The final dataset contains $513\,303$ submissions by $157\,528$ unique users who contribute a mean (median) of $2.57$ ($1$) submissions.

\xhdr{Calorie distribution over the years}
To estimate the macro-nutrient content of foods shared on Reddit, we applied our calorie approximation approach to $513,303$ posts on the popular r/food subreddit (Fig.~\ref{fig:macro_nutrients}).
Our analysis reveals seasonal patterns, such as an increase in calorie levels toward the end of each year—a trend likely attributable to users sharing more indulgent foods during festive occasions such as Christmas (Fig.~\ref{fig:macro_nutrients}, left panel).
Most notably, our estimations indicate a substantial increase in calories between March and June 2020, coinciding with the early phase of the COVID-19 pandemic. This plateau of higher calorie values suggests a temporary shift in the types of foods people shared online, further underscored by a rise in carbohydrate-rich foods during the same period (Fig.~\ref{fig:macro_nutrients}, right panel).
While these trends and patterns serve to demonstrate the insights enabled by our approach, future research could investigate these phenomena in greater detail to better understand food-sharing behavior.
Considering online sharing behavior as a potential proxy for real-world eating habits, this analysis holds substantial promise for uncovering consumption patterns across a large population of users.
Such findings could have significant implications for both computational social science and health research.

\section{Conclusion}
In this work, we present a novel method for estimating macro-nutrients from food titles.
By combining the USDA food database with SentenceTransformer embeddings, our approach demonstrates strong performance on a ground-truth dataset of online recipe titles and competes effectively with industrial calorie estimation APIs. 
Furthermore, we apply our method to over $500\,000$ real food posts on Reddit, enabling a longitudinal analysis of global food trends on this popular online platform.
Overall, our method and accompanying code offer a flexible tool for researchers investigating food-sharing practices or eating behaviors across various online domains.

\bibliographystyle{ACM-Reference-Format}
\bibliography{ref}


\begin{thebibliography}{9}

%%% the Web address.

\ifx \showCODEN    \undefined \def \showCODEN     #1{\unskip}     \fi
\ifx \showDOI      \undefined \def \showDOI       #1{#1}\fi
\ifx \showISBNx    \undefined \def \showISBNx     #1{\unskip}     \fi
\ifx \showISBNxiii \undefined \def \showISBNxiii  #1{\unskip}     \fi
\ifx \showISSN     \undefined \def \showISSN      #1{\unskip}     \fi
\ifx \showLCCN     \undefined \def \showLCCN      #1{\unskip}     \fi
\ifx \shownote     \undefined \def \shownote      #1{#1}          \fi
\ifx \showarticletitle \undefined \def \showarticletitle #1{#1}   \fi
\ifx \showURL      \undefined \def \showURL       {\relax}        \fi
% invisible to TeX
\providecommand\bibfield[2]{#2}
\providecommand\bibinfo[2]{#2}
\providecommand\natexlab[1]{#1}
\providecommand\showeprint[2][]{arXiv:#2}

\bibitem[Baumgartner et~al\mbox{.}(2020)]%
        {baumgartner2020pushshift}
\bibfield{author}{\bibinfo{person}{Jason Baumgartner}, \bibinfo{person}{Savvas Zannettou}, \bibinfo{person}{Brian Keegan}, \bibinfo{person}{Megan Squire}, {and} \bibinfo{person}{Jeremy Blackburn}.} \bibinfo{year}{2020}\natexlab{}.
\newblock \showarticletitle{The pushshift reddit dataset}. In \bibinfo{booktitle}{\emph{Proceedings of the international AAAI conference on web and social media}}, Vol.~\bibinfo{volume}{14}. \bibinfo{pages}{830--839}.
\newblock


\bibitem[De~Choudhury et~al\mbox{.}(2016)]%
        {de2016characterizing}
\bibfield{author}{\bibinfo{person}{Munmun De~Choudhury}, \bibinfo{person}{Sanket Sharma}, {and} \bibinfo{person}{Emre Kiciman}.} \bibinfo{year}{2016}\natexlab{}.
\newblock \showarticletitle{Characterizing dietary choices, nutrition, and language in food deserts via social media}. In \bibinfo{booktitle}{\emph{Proceedings of the 19th acm conference on computer-supported cooperative work \& social computing}}. \bibinfo{pages}{1157--1170}.
\newblock


\bibitem[Fukagawa et~al\mbox{.}(2022)]%
        {fukagawa2022usda}
\bibfield{author}{\bibinfo{person}{Naomi~K Fukagawa}, \bibinfo{person}{Kyle McKillop}, \bibinfo{person}{Pamela~R Pehrsson}, \bibinfo{person}{Alanna Moshfegh}, \bibinfo{person}{James Harnly}, {and} \bibinfo{person}{John Finley}.} \bibinfo{year}{2022}\natexlab{}.
\newblock \showarticletitle{USDA's FoodData Central: what is it and why is it needed today?}
\newblock \bibinfo{journal}{\emph{The American journal of clinical nutrition}} \bibinfo{volume}{115}, \bibinfo{number}{3} (\bibinfo{year}{2022}), \bibinfo{pages}{619--624}.
\newblock


\bibitem[Gligori{\'c} et~al\mbox{.}(2020)]%
        {gligoric2020experts}
\bibfield{author}{\bibinfo{person}{Kristina Gligori{\'c}}, \bibinfo{person}{Manoel~Horta Ribeiro}, \bibinfo{person}{Martin M{\"u}ller}, \bibinfo{person}{Olesia Altunina}, \bibinfo{person}{Maxime Peyrard}, \bibinfo{person}{Marcel Salath{\'e}}, \bibinfo{person}{Giovanni Colavizza}, {and} \bibinfo{person}{Robert West}.} \bibinfo{year}{2020}\natexlab{}.
\newblock \showarticletitle{Experts and authorities receive disproportionate attention on Twitter during the COVID-19 crisis}.
\newblock \bibinfo{journal}{\emph{arXiv}} (\bibinfo{year}{2020}).
\newblock


\bibitem[Lazer et~al\mbox{.}(2021)]%
        {lazer2021meaningful}
\bibfield{author}{\bibinfo{person}{David Lazer}, \bibinfo{person}{Eszter Hargittai}, \bibinfo{person}{Deen Freelon}, \bibinfo{person}{Sandra Gonzalez-Bailon}, \bibinfo{person}{Kevin Munger}, \bibinfo{person}{Katherine Ognyanova}, {and} \bibinfo{person}{Jason Radford}.} \bibinfo{year}{2021}\natexlab{}.
\newblock \showarticletitle{Meaningful measures of human society in the twenty-first century}.
\newblock \bibinfo{journal}{\emph{Nature}} \bibinfo{volume}{595}, \bibinfo{number}{7866} (\bibinfo{year}{2021}), \bibinfo{pages}{189--196}.
\newblock


\bibitem[Reimers and Gurevych(2019)]%
        {reimers2019sbert}
\bibfield{author}{\bibinfo{person}{Nils Reimers} {and} \bibinfo{person}{Iryna Gurevych}.} \bibinfo{year}{2019}\natexlab{}.
\newblock \showarticletitle{Sentence-BERT: Sentence Embeddings using Siamese BERT-Networks}. In \bibinfo{booktitle}{\emph{Proceedings of the 2019 Conference on Empirical Methods in Natural Language Processing}}. \bibinfo{publisher}{Association for Computational Linguistics}.
\newblock
\urldef\tempurl%
\url{https://arxiv.org/abs/1908.10084}
\showURL{%
\tempurl}


\bibitem[Ruprechter et~al\mbox{.}(2023)]%
        {ruprechter2023poor}
\bibfield{author}{\bibinfo{person}{Thorsten Ruprechter}, \bibinfo{person}{Keith Burghardt}, {and} \bibinfo{person}{Denis Helic}.} \bibinfo{year}{2023}\natexlab{}.
\newblock \showarticletitle{Poor attention: The wealth and regional gaps in event attention and coverage on Wikipedia}.
\newblock \bibinfo{journal}{\emph{PLoS one}} \bibinfo{volume}{18}, \bibinfo{number}{11} (\bibinfo{year}{2023}), \bibinfo{pages}{e0289325}.
\newblock


\bibitem[Russo et~al\mbox{.}(2024)]%
        {russo2024stranger}
\bibfield{author}{\bibinfo{person}{Giuseppe Russo}, \bibinfo{person}{Manoel~Horta Ribeiro}, {and} \bibinfo{person}{Robert West}.} \bibinfo{year}{2024}\natexlab{}.
\newblock \showarticletitle{Stranger Danger! Cross-Community Interactions with Fringe Users Increase the Growth of Fringe Communities on Reddit}. In \bibinfo{booktitle}{\emph{Proceedings of the International AAAI Conference on Web and Social Media}}, Vol.~\bibinfo{volume}{18}. \bibinfo{pages}{1342--1353}.
\newblock


\bibitem[West et~al\mbox{.}(2013)]%
        {west2013cookies}
\bibfield{author}{\bibinfo{person}{Robert West}, \bibinfo{person}{Ryen~W White}, {and} \bibinfo{person}{Eric Horvitz}.} \bibinfo{year}{2013}\natexlab{}.
\newblock \showarticletitle{From cookies to cooks: Insights on dietary patterns via analysis of web usage logs}. In \bibinfo{booktitle}{\emph{Proceedings of the 22nd international conference on World Wide Web}}. \bibinfo{pages}{1399--1410}.
\newblock


\end{thebibliography}

\end{document}